\documentclass[%
 reprint,
nofootinbib,
 amsmath,amssymb,
aps,
floatfix,
]{revtex4-2}

\usepackage{graphicx}%
\usepackage{dcolumn}%
\usepackage{bm}%

\usepackage{xspace}
\usepackage[babel]{csquotes}
\usepackage[list-units=single]{siunitx}
\usepackage{physics}
\usepackage[colorlinks,linkcolor={blue!70!black},citecolor={blue!70!black},urlcolor={blue!70!black}]{hyperref}
\usepackage[noabbrev,nameinlink]{cleveref}
\usepackage[acronym]{glossaries}\glsdisablehyper
\usepackage{orcidlink}
\usepackage{ifthen}
\usepackage{placeins}
\usepackage{float}

\newcommand\mean[1]{\ensuremath{\left\langle{#1}\right\rangle}}
\newcommand{\naturals}{\ensuremath{\mathbb{N}}}
\newcommand{\x}{\ensuremath{\text{x}}\xspace}
\newcommand{\ex}{\ensuremath{\text{ex}}\xspace}
\newcommand{\rev}{\ensuremath{\text{rev}}\xspace}
\newcommand{\res}{\ensuremath{\text{res}}\xspace}

\newcommand{\dtc}{\ensuremath{{\Delta t_\text{count}}}\xspace}
\newcommand{\dte}{\ensuremath{{\Delta t_\text{eval}}}\xspace}

\newcommand{\qex}[1][]{\ensuremath{Q_{\ex\ifthenelse{\equal{#1}{}}{}{,#1}}}\xspace}
\newcommand{\dqex}[1][]{\ensuremath{\Delta\qex[#1]}\xspace}

\newcommand{\ion}[3]{\ensuremath{{}^{#1}\text{#2}{}^{#3}}}

\newcommand\Npp{Noise\nolinebreak[4]\,\raisebox{.08ex}{\small{{++}}}\xspace}

\newcommand{\changed}[1]{#1\xspace}
\newcommand{\changedd}[1]{#1\xspace}

\newacronym{am}{AM}{amplitude modulation}
\newacronym{btf}{BTF}{beam transfer function}
\newacronym{fm}{FM}{frequency modulation}
\newacronym{gsi}{GSI}{GSI Helmholtz Centre for Heavy Ion Research}
\newacronym{rf}{RF}{radio frequency}
\newacronym{rfko}{RF-KO}{Radio Frequency Knock Out} %
\newacronym{sis18}{SIS18}{Heavy Ion Synchrotron}

\newcommand{\sis}{\ifglsused{sis18}{}{the }\gls{sis18} at \gls{gsi}\xspace}

\graphicspath{{images}{.}}\glsunset{gsi}

\begin{document}

\title{
    Excitation of nonlinear second order betatron sidebands\\
    for Knock Out slow extraction at the \changed{third-integer} resonance
}

\author{Philipp Niedermayer\,\orcidlink{0000-0002-4722-6619}}
\email{p.niedermayer@gsi.de}
\author{Rahul Singh\,\orcidlink{0000-0002-7427-8000}}
\affiliation{GSI Helmholtzzentrum für Schwerionenforschung, Darmstadt, Germany}

\date{\today}

\begin{abstract}
\Glsentrylong{rfko} resonant slow extraction is a standard method for extracting stored particle beams from synchrotrons by transverse excitation.
Classically, the beam is excited with an RF field comprising a frequency band around one of the betatron sidebands.
This article demonstrates that the \changed{third-integer} resonance commonly used for the slow extraction induces nonlinear motion, resulting in the appearance of additional sidebands of higher order at multiples of the betatron tune.
Measured and simulated beam spectra are presented, revealing these sidebands and the beam's response to being excited at first and second order sidebands.
The feasibility of using a second order sideband for the purpose of slow extraction is demonstrated.
This results in a significant improvement in the temporal structure (spill quality) of the extracted beam, but at the cost of higher excitation power requirements.
This is observed both experimentally and in tracking simulations.
The mechanism behind the observed improvement is explained using beam dynamics simulations.
\end{abstract}

\maketitle

\section{Introduction}

\Gls{rfko} resonant slow extraction is used to extract stored particle beams from synchrotron rings \cite{Hiramoto.1992,Tomizawa.1993}.
Therefore, a beam optics near a \changed{third-integer} resonance is chosen where sextupole fields create a transverse, betatron amplitude dependent instability (separatrix) \cite{PIMMS.1999}.
For storing and accelerating a beam such instabilities are undesired and investigated to be corrected \cite{Franchi.2014};
but in the context of slow extraction they are exploited and enhanced by dedicated sextupole magnets.
The beam is driven into the instability in a controlled manner by increasing the particle amplitude through transverse excitation.
The excitation system consists of \ifglsused{rf}{an}{a} \gls{rf} signal generator, \gls{rf} amplifiers, and a {stripline kicker} \cite{BelverAguilar.2014}, inside of which electromagnetic \gls{rf} fields deflect traversing particles on each turn.
When particles reach the separatrix, their motion becomes unbound and the betatron amplitude increases rapidly, such that septa can be used to deflect the extracted particles into an extraction beam line.
This \emph{spill} of extracted particles is then delivered to experiments or \changed{used for} medical therapy.

\subsection{Nonlinear betatron oscillation}
\label{sec:nonlinear_dynamics}

With the working point (\changed{betatron} tune) of the circular accelerator close to a \changed{third-integer} resonance
driven by nonlinear sextupole fields, %
the three-turn particle dynamics is described by the Kobayashi Hamiltonian \cite{Kobayashi.1970}
\begin{align*}
    H &= 3\pi d \left(X^2 + X'^2\right) + \frac{S}{4} \left(3 X X'^2 - X^3\right)
    \\&= 6\pi d J_\x - \frac{S}{\sqrt{2}} J_\x^{3/2} \cos(3\Theta_\x)
\end{align*}
where $S = -\beta_x(s)^{3/2} k_2 l / 2$ is the normalized sextupole strength %
and
\begin{align}
\label{eq:phase_space_coordinates}
    X  &= \sqrt{2J_\x} \cos(\Theta_\x)
       = \frac{1}{\sqrt{\beta_\x(s)}} \, x(s)
    \\
    X' &= -\sqrt{2J_\x} \sin(\Theta_\x)
       = -\frac{\beta^\prime_\x(s)}{2\sqrt{\beta_\x(s)}} \, x(s) 
         +\sqrt{\beta_\x(s)} \,  x'(s)
\nonumber
\end{align}
are the normalized phase space coordinates with the action~$J_\x$ and angle~$\Theta_\x$ \cite{PIMMS.1999}.
The relation to the physical coordinate\footnote{Corrected for closed orbit and dispersion}~$x$ and divergence~$x'$ is thereby given by the beta function~$\beta_\x(s)$ which describes the optical properties of the lattice at the location $s$.
The (small) quantity $d = Q_\x - Q_\res$ is the distance of the tune\footnote{Including chromatic detuning: $Q_\x = Q_{\x,\text{ref}} + \xi_\x \delta$}~$Q_\x$ to the \changed{third-integer} resonance~$Q_\res$ with $3Q_\res\in\naturals$.
In the following, uppercase symbols will be used to denote absolute tune values, while the lowercase $q \leq 0.5$ refers to the fractional tune as distance to the nearest integer. %
\changedd{At the separatrix, the Kobayashi Hamiltonian reaches a value of $H_\text{sep} = {\left(4\pi d\right)^3}/{S^2}$.}
\\

\label{sec:detuning}

In the linear case ($S=0$) %
particles perform harmonic betatron oscillations with a constant phase advance per turn $\mu_1 = 2\pi Q_\x$. %
Near the resonance, however, the betatron oscillations becomes nonlinear and the sextupole field causes an amplitude and phase dependent detuning.
In \changed{the} thin lens and flat beam approximation\changed{s}, the kick induced by the sextupole is $\Delta X' = S X^2$.
For the angle after the sextupole one can thus write
\begin{equation*}
	\tan(\Theta_{\x,1})
	= -\frac{X' + S X^2}{X}
	= \tan(\Theta_\x) - S\sqrt{2J_\x}\cos(\Theta_\x)
\end{equation*}
Considering that the sextupole kick is small and using the small angle approximation
\begin{equation*}
    \tan(\Theta_\x+\Delta\Theta_\x)
	\approx \tan(\Theta_\x) + \frac{\Delta\Theta_\x}{\cos^2(\Theta_\x)} + \order{\Delta\Theta_\x^2}
\end{equation*}
the change in angle follows to first order as
\begin{equation*}
    \Delta\Theta_\x 
    = \Theta_{\x,1} - \Theta_\x
    = - S\sqrt{2J_\x}\cos^3(\Theta_\x)
    \ll 1
\end{equation*}
As a result, for the nonlinear betatron oscillation, the phase advance per turn
\begin{equation}
\label{eq:phase_advance_detuned}
    \tilde{\mu}_1
	= 2 \pi Q_\x - S\sqrt{2J_\x}\cos^3(\Theta_\x)\\
\end{equation}
shows an amplitude and phase dependent detuning.

As a particle with constant $H$ revolves in phase space, the detuning term effectively modulates the betatron frequency as depicted in \cref{fig:detuning}.
This frequency modulation generates anharmonic motion and sidebands of higher order.

\begin{figure}[t]
    \centering
    \includegraphics[scale=0.75]{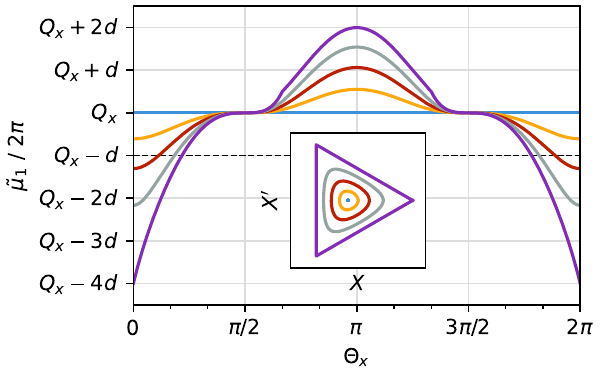}
    \caption{%
        Nonlinear detuning of the phase advance per turn as function of angle,
        calculated \changed{with \cref{eq:phase_advance_detuned}} for
        \changed{constant $H=0$ (blue) to $H=H_\text{sep}$ (purple)}. %
        The dashed line marks the resonance condition.
        The inset shows the corresponding phase space trajectories
        \changed{calculated with \cref{eq:phase_space_coordinates}}.
    }
    \label{fig:detuning}
\end{figure}

\section{Nonlinear beam spectra \\under excitation}

Spectra of transverse beam motion are obtained by observing the position of the circulating beam as a function of time at a fixed location in the synchrotron.
For the linear case, the resulting spectra are well understood and dominated by the betatron sidebands at $f = (n \pm q_\x) f_\rev$ below and above each harmonic $n \in \naturals$ of the revolution frequency $f_\rev$.
These harmonics are related to the transverse position of the beam centroid \cite{Singh.2018b}.
The sidebands are the result of the amplitude modulation of this position due to betatron oscillations and are used to determine linear quantities like tune and chromaticity \cite{Betz.2017}.

In the nonlinear case the betatron oscillation are frequency modulated.
Therefore, one expects multiple sidebands of order $k \in \naturals$ around each harmonic $n$:
\begin{equation*}
    {f} = \left( n \pm k \, q_\x \right) {f_\rev}
\end{equation*}

\subsection{Measurement of transfer function}

A \gls{btf} measurement is performed by exciting the beam with a fixed frequency $f_\ex$ and observing the magnitude and phase of the induced beam oscillations filtered at that specific excitation frequency. %
The measurement is repeated within a single machine cycle for a range of frequencies, yielding the beam response as function of frequency.
For a linear system the transfer function is well defined as the complex quotient of the observed and induced oscillation \cite{Borer.1979}.
However, in the nonlinear case \changed{the amplitude detuning affects the measurement by altering the} oscillation frequencies present in the excited system, \changed{as the excitation causes particles to be depleted from the beam core to higher amplitudes, where they are detuned.
Since the \gls{btf} is solely sensitive to oscillations at the excitation frequency, it can thus only represent a subset of the frequencies and nonlinear dynamics involved
and becomes sensitive to the particle distribution and the speed and amplitude of the excitation frequency sweep}
\cite{Cortes.20XX}. %

Nevertheless, a \gls{btf} measurement can be used to probe the beam response at the frequency where the nonlinear sidebands of second order are expected.
Such a measurement is presented in \cref{fig:exp:btf} for a typical \gls{rfko} extraction setup from \sis
\changed{using a vector network analyzer connected to a stipline kicker and pickup}.
Alongside the nonlinear case with sextupoles, also the linear case with sextupole magnets switched off is shown.
As the closed orbit was not corrected, the sextupole magnets cause a tune shift.
Therefore the location of the first and second order sidebands is indicated for both cases as obtained by fitting of the respective phase response.

\begin{figure}[!b]
    \centering
    \includegraphics[scale=0.75]{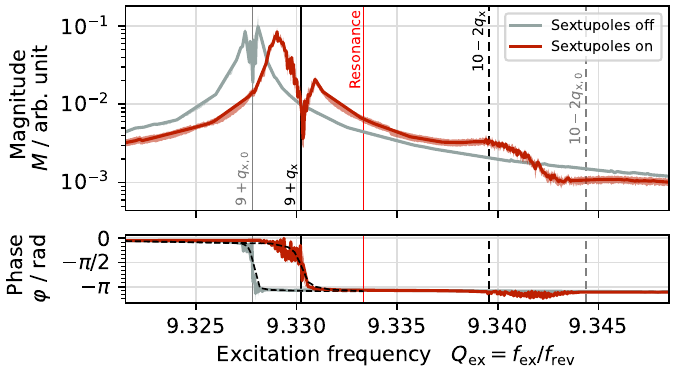}
    \caption{%
        Measured response
        to a sinusoidal excitation of varying frequency (BTF measurement) %
        of an \ion{197}{Au}{65+} beam at \SI{800}{MeV/nucleon}.
        Vertical lines indicate the location of the betatron sidebands and \changed{third-integer} resonance.
    }
    \label{fig:exp:btf}
\end{figure}

\begin{figure*}
    \centering
    \includegraphics[scale=0.74]{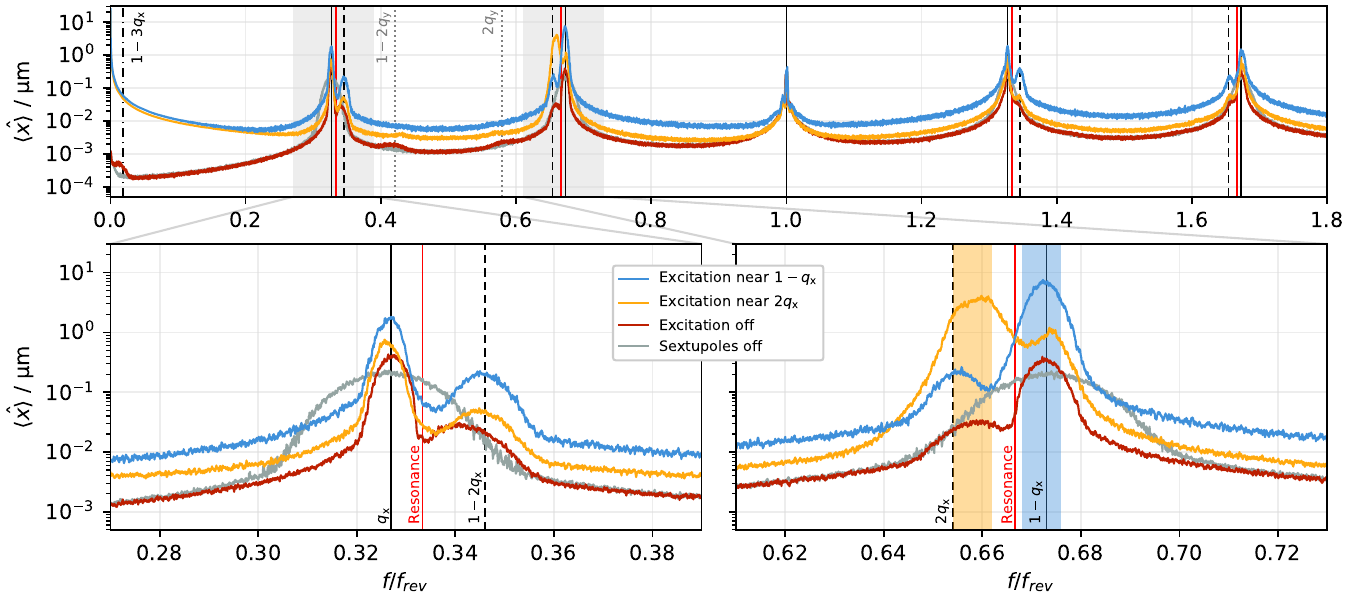}
    \caption{%
        Transverse Schottky spectra for the linear case (sextupoles off) and three nonlinear cases (sextupoles on) with and without excitation as obtained from a simulation.
        The zoom parts show beam oscillations at first and second order sidebands (vertical black lines) and at the excitation bands (yellow and blue shaded regions).
    }
    \label{fig:sim:schottky_spectra}
\end{figure*}

A response of the beam to an excitation near the second order betatron sideband \changed{at $10-2q_\x$} can only be observed for the nonlinear case with sextupoles switched on.
\changed{The response is measured at the $10^\text{th}$ harmonic because of the higher transfer impedance of the \SI{50}{\ohm} terminated pickup.}
The measurement demonstrates the existence of the second order sideband and the feasibility of exciting it.
While the response is very weak compared to the first order sideband, one has to keep in mind that the \gls{btf} does not give insight into the beam oscillations at any but the excited frequency.
For example, spectral components near the second order sideband induced by excitation at the first order sideband are not visible.
However, the dip in the response at the first order sideband suggest, that in this case the primary oscillation frequencies are shifted away from the excitation frequency where the \gls{btf} is not sensitive \cite{Cortes.20XX}.  %
Therefore, in the following, the beam motion is investigated by means of Schottky spectra in simulations.

\subsection{Simulation of Schottky spectra}

Schottky diagnostics is a vital tool to study the incoherent particle motion by sampling the beam position at frequencies higher than the revolution frequency \changed{\cite{Boussard.1995}}.
Unlike the \gls{btf}, it provides a complete picture of the intrinsic beam oscillations independent of an external excitation.
A particle tracking simulation is performed with Xsuite \cite{Iadarola.2023}
for a typical machine setup of a Knock Out extraction from \sis.
A beam of \ion{238}{U}{73+} ions with
a rigidity of $B\rho = \SI{7}{\tesla\meter}$
and revolution frequency of $f_\rev=\SI{785}{kHz}$
is prepared for extraction using a machine tune of $Q_{\x,\text{ref}} = 4.327$ and a corrected orbit.

\Cref{fig:sim:schottky_spectra} shows the spectra of the horizontal beam centroid $\mean{x}$ as obtained from the simulation using a sampling frequency of $\SI{3.92}{MHz} = 5f_\rev$.
In the linear case (sextupoles off), the regular betatron sidebands of first order are visible, which are relatively broad due to the large chromaticity of $\xi_\x = \Delta Q_\x / \delta = -6.6$.
When the sextupoles are switched on for the purpose of slow extraction, %
the chromaticity reduces to $\xi_\x = -2.0$ and the first order sidebands become narrower. %
In addition, the sidebands of higher order ($k \ge 2$) become visible, most prominently the second order sidebands at $n \pm 2 q_\x$.
But also the weaker third order sideband at $1-3q_\x = 3 \abs{d} \approx 0$ and faint coupling bands at $n \pm 2 q_\text{y}$ can be observed.
The latter result from the fact that the sextupole fields also depend on the vertical coordinate, thus coupling the betatron motion in both planes.
\changedd{%
The observation that the emergence of second and higher order sidebands is related to the nonlinear dynamics discussed in \cref{sec:nonlinear_dynamics} is further emphasized in \cref{fig:sim:schottky_hollow_beam}.
Here, Schottky spectra of beam slices for discrete ranges of the Hamiltonian from the core to the separatrix are depicted (compare also \cref{fig:detuning}).
The plot shows not only how the frequency of the sidebands shifts due to the amplitude detuning,
but also reveals a clear correlation of the particles' Hamiltonian with the strength of the nonlinear second and third order sidebands.
An external excitation at the second order sideband will thus couple strongly to those particles close to the separatrix undergoing strong nonlinear motion, while only weakly affecting particles in the core.
}
\\

\begin{figure*}
    \centering
    \includegraphics[width=\linewidth]{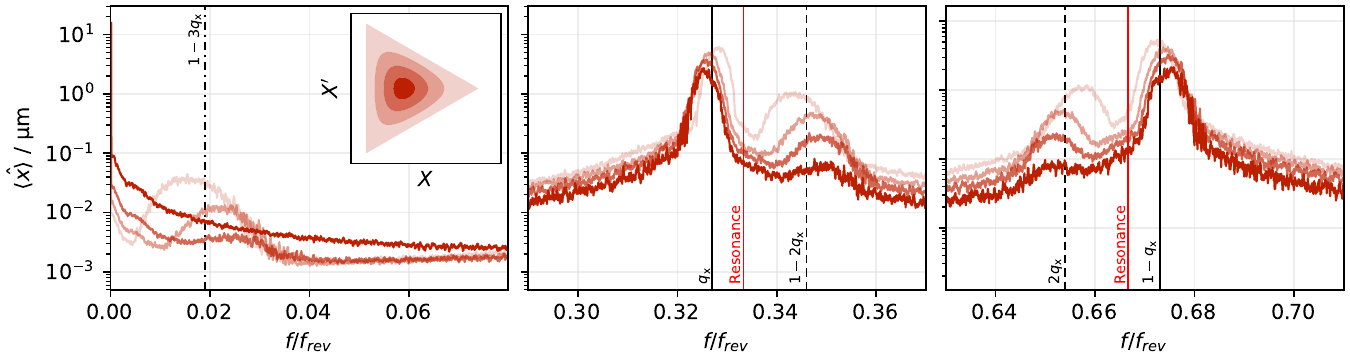}
    \caption{\changedd{%
        Transverse Schottky spectra (sextupoles on, excitation off)
        as function of phase space amplitude by means of the Kobayashi Hamiltonian $H$ as obtained from a simulation.
        The shading corresponds to four beam slices of particles 
        in the range $h_i < H/H_\text{sep} < h_{i+1}$ 
        with $h_i = (i/4)^2$ and $i \in \naturals$.
        The inset plot shows these ranges in normalized phase space.
    }}
    \label{fig:sim:schottky_hollow_beam}
\end{figure*}

In \gls{rfko} extraction, the beam is excited transversely with a signal of a certain bandwidth \changed{to account for its intrinsic tune spread.
While many different signal types can be used depending on the application, a band-limited noise signal is used here to avoid the bias of any specific non-uniform excitation spectrum}.
Under the excitation, the betatron oscillation amplitude is increased, and so does the magnitude in the respective Schottky spectra in \cref{fig:sim:schottky_spectra}.
When the amplitude grows beyond the separatrix where particles are extracted, a significant DC leakage into the low frequencies is observed in the Schottky spectra.
This is due to the slow variation in centroid position compared to the case without extraction, where the mean beam position remains at zero.

Traditionally, the excitation \changed{band is placed} in the vicinity of one of the first order betatron sidebands.
\changed{This is the consequence of modelling the excitation as a forced oscillation of individual particles undergoing simple harmonic motion under linear restoring force.}  %
\changed{Motivated by the nonlinear dynamics discussed above, a second case is considered where the excitation band is placed near the second order betatron sideband.
The choice of two nearby sidebands around $f/f_\rev \approx 1-q_\x \approx 2q_\x \approx 2/3$ avoids a systematic bias on the spill quality stemming from a large difference of excitation frequencies.
It also allows to distinguish the behaviour of the excited bands from the corresponding basebands around $f/f_\rev \approx 1/3$.
As can be seen in \cref{fig:sim:schottky_spectra}, in both cases} the excitation induces coherent beam oscillations only at the \emph{absolute} excitation frequency, \changed{whereas the spectral response at any other betatron sideband at the other harmonics remains one order of magnitude weaker.}
This becomes especially obvious for the excitation band placed near $2q_\x$, for which the oscillation \changed{magnitude} of the excited second order sideband is increased even beyond the \changed{magnitude} of the nearby first order sideband $1-q_\x$.
Such a behaviour is only observed at the excited sideband, and not at any of the other sidebands, which are not directly excited.
In general, the \changed{overall} magnitudes observed in the Schottky spectrum for the case of excitation near the second order sideband \changed{$2q_\x$} are smaller compared to the case of excitation at the first order sideband \changed{$1-q_\x$, even though in the former case the amplitude of the excitation signal used in the simulation is larger to achieve an equal extraction rate}.
This suggests that the coupling to the beam is weaker at the higher order sidebands, but it is still sufficient for driving \ifglsused{rf}{an}{a} \gls{rfko} extraction, \changed{as demonstrated in the upcoming section}.

\section{Knock Out extraction with second order sidebands} %

To study the effect of an excitation at a second order sideband on the \gls{rfko} slow extraction process, the required excitation power and achievable spill quality is investigated.
The spill quality is given by the amount of unwanted intensity fluctuations in the time structure of the extracted beam.
To quantify the fluctuations, the spill is divided into time intervals of length \dtc and the number of extracted particles $N$ is counted in each interval.
The standard deviation $\sigma=\sqrt{\mean{(N-\mean{N})^2}}$ and mean $\mu=\mean{N}$, evaluated over a larger time span \dte, is then used to determine
the coefficient of variation~$c_\text{v}$
or the equivalent spill duty factor~$F$
\cite{Singh.2018}:
\begin{align*}
	c_\text{v}
	=\frac{\sigma}{\mu}
	=\sqrt{\frac{\mean{N^2}}{\mean{N}^2} - 1}
&\quad&
	F
	= \frac{\mean{N}^2}{\mean{N^2}}
	= \frac{1}{1 + c_\text{v}^2}
\end{align*}

In addition to the spill quality, also the extraction efficiency is of importance.
To extract all particles from the storage ring, a certain excitation power is required which depends strongly on the beam rigidity and the excitation signal used.
In practice the available power of the \gls{rf} amplification system can impose limits on the excitation signals which are feasible.
\\

\subsection{Machine measurement}

\begin{figure*}
    \centering
    \includegraphics[scale=0.75]{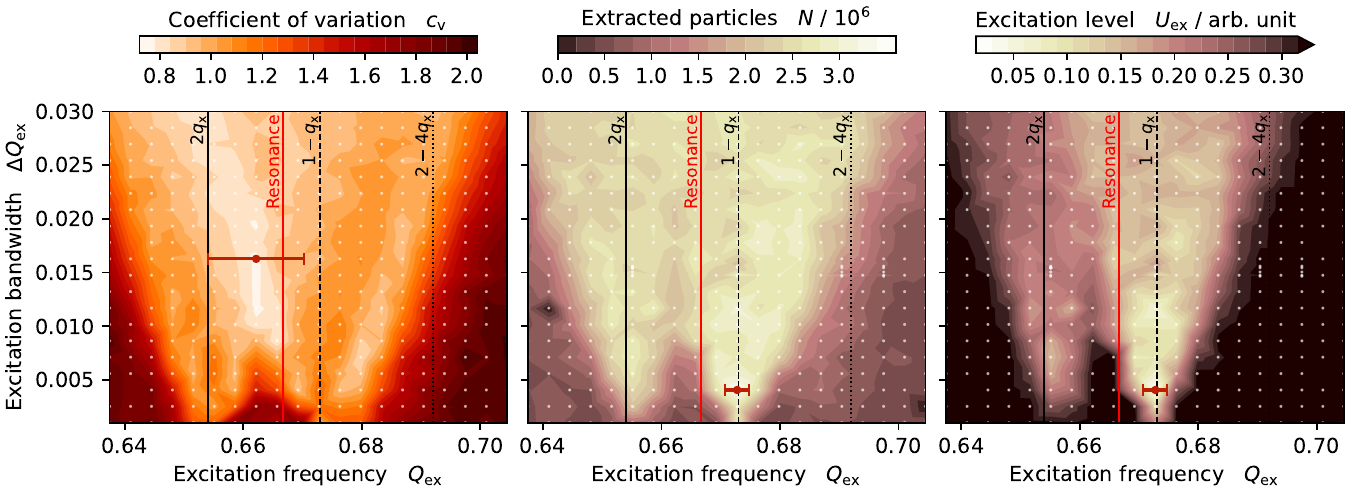}
    \caption{%
        Measured spill quality (left),
        number of particles extracted (middle)
        and required excitation strength (right)
        as a function of central frequency and bandwidth of a band-limited noise excitation signal.
        The coefficient of variation is calculated for $\dtc=\SI{100}{\us}$ and $\dte=\SI{1}{s}$.
        \changed{The coloured contour plots are derived by triangulation from the sampled points (white dots).} %
        The red dot and span marks the respective global optimum and associated bandwidth.
        The vertical lines indicate the \changed{third-integer} resonance and nearby betatron sidebands.
    }
    \label{fig:exp:scan_qex_noise}
\end{figure*}

The effects of excitation frequency and bandwidth on the \gls{rfko} extraction are studied experimentally at the \gls{sis18} of \gls{gsi}.
An \ion{238}{U}{73+} beam with a kinetic energy of \SI{200}{MeV/nucleon} (%
$B\rho = \SI{7}{\tesla\meter}$, 
$f_\rev=\SI{785}{kHz}$%
) is prepared and excited with a band-limited noise signal.
The extracted spill is observed with a \changed{particle detector based on a plastic scintillator} \cite{Forck.1997}, which serves two purposes:
First, the signal is used with a feedback system to dynamically adjust the excitation signal amplitude over the duration of the spill in order to maintain a constant spill rate of \SI{e6}{particles/s} \cite{Niedermayer.2023a}.
Secondly, the intensity fluctuations measured by the detector are evaluated to determine the spill quality.
\changed{Thereby $\dtc=\SI{100}{\us} \gg f_\text{ex}^{-1}$ was chosen to provide sufficient counting statistics for the quality analysis.} %

\Cref{fig:exp:scan_qex_noise} shows the results of a large scan of the excitation frequency and bandwidth, exceeding the traditionally used vicinity of the first order betatron sideband.
The measurement proves, that \gls{rfko} slow extraction by exciting the second order betatron sideband is possible, and a comparable number of particles can be extracted, namely
$\mean{N}_{2q_\x}/\mean{N}_{1-q_\x} = \num{2.4+-0.3} / \num{2.7+-0.3} = \SI{90}{\percent}$. %
The excitation level required is with
$\mean{U_\ex}_{2q_\x}/\mean{U_\ex}_{1-q_\x} = \num{0.21+-0.04} / \num{0.12+-0.05} = \num{1.7+-0.6}$ %
about twice as large as for the first order sideband.
The optimal choice of excitation frequency and bandwidth from an efficiency point of view hence lies in the close vicinity of the first order betatron sideband.

However, the optimum in terms of spill quality was found on the opposite side of the resonance half way towards the second order betatron sideband (\cref{fig:exp:scan_qex_noise}, left).
Here, the measurement shows a reduction of the intensity fluctuations to $c_\text{v}=\num{0.76}$ %
compared to $c_\text{v}=\num{1.21}$ %
for the efficiency optimum at the first order sideband.
This improvement in spill quality comes with an increased power requirement.
While the transverse \gls{rf} excitation system is capable of delivering these demands for the presented conditions, in general for higher rigidity beams a trade-off between spill quality and required power has to be found.

\subsection{Simulation of spill quality}

\begin{figure}[b]
    \centering
    \includegraphics[scale=0.75]{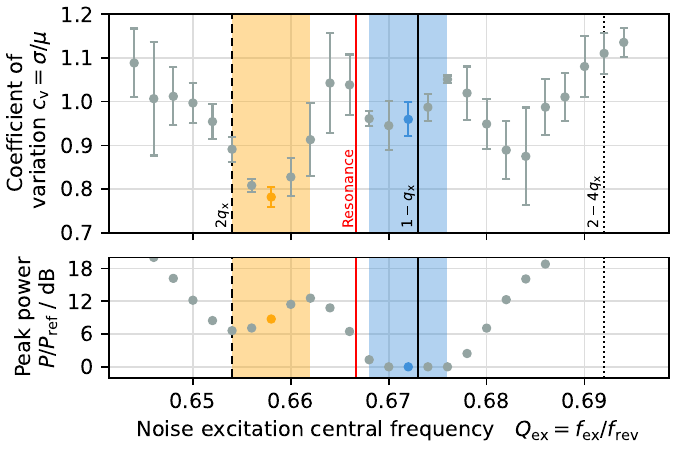}
    \caption{%
        Spill quality (top) and excitation power (bottom) as function of excitation frequency for a band-limited noise signal ($\dqex=\num{0.008}$) as obtained from simulations.
        The quality is calculated for $\dtc=\SI{100}{\us}$ and $\dte=\SI{100}{\ms}$.
        Vertical lines mark the resonance and nearby betatron sidebands, and the two excitation bands from \cref{fig:sim:schottky_spectra} are indicated.
    }
    \label{fig:sim:scan_qex_noise_medium}
\end{figure}

\begin{figure*}
    \centering
    \includegraphics[scale=0.75]{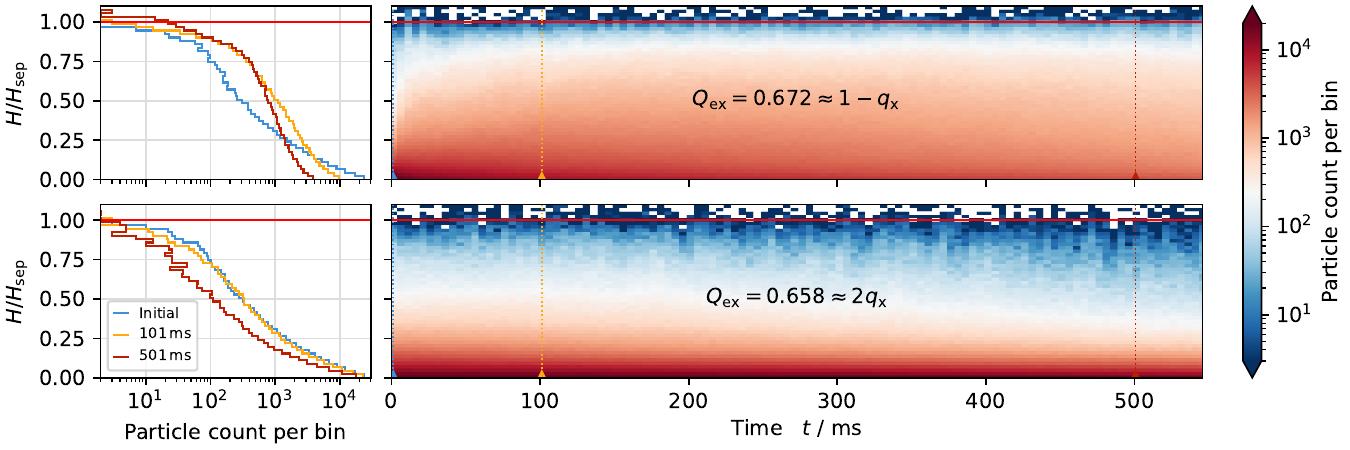}
    \caption{%
        Particle distributions before and during excitation near the first (top) and second order sideband (bottom)
        obtained from a simulation of $10^5$ particles.
        The band-limited noise excitation signals correspond to \cref{fig:sim:schottky_spectra,fig:sim:scan_qex_noise_medium} ($\dqex=\num{0.008}$).
        The location of the separatrix at $H=H_\text{sep}$ is indicated by the solid red line.
    }
    \label{fig:sim:beam_dynamics_excited}
\end{figure*}

The previously introduced simulation code is used to model the machine experiment.
\Cref{fig:sim:scan_qex_noise_medium} shows the spill quality and required excitation power as a function of the excitation frequency as obtained from the simulation.
Thereby, the bandwidth of the band-limited noise signal remains fixed.
In agreement with the measurement, the global optimum in terms of spill quality is located on the opposite side of the resonance towards the second order betatron sideband.
While the spill fluctuations are significantly reduced compared to an excitation band near the first order betatron sideband, the required excitation power is about \SI{9}{dB} higher, corresponding to a factor three larger signal amplitude.
The increase in the required excitation power is consistent with the experimental findings and the observations from the Schottky spectra discussed in the previous section.

Another local optimum can be found towards the fourth order betatron sideband $2-4q_\x$, where the excitation power required is even higher.
Once the power limit is reached, only a small fraction of the beam can be extracted and the value of $c_\text{v}\propto 1/\mu$ increases again.

\subsection{Simulation of beam dynamics} %

To understand the mechanism behind the observed spill quality improvement, %
the beam dynamics is analysed in the particle tracking simulations.
For the two excitation bands near the first and second order betatron sidebands discussed above, the particle distribution in horizontal phase space is recorded as a function of time.
The Kobayashi Hamiltonian $H$ normalized to its value at the separatrix $H_\text{sep}$ is used \changed{to analyse} the particle distribution during the extraction \changed{by means of a histogram} (\cref{fig:sim:beam_dynamics_excited}).
Thereby, $H=0$ corresponds to a particle in the centre of phase space with zero betatron amplitude and (chromatic) tune $Q_\x$;
and for $H \to H_\text{sep}$ the betatron amplitude and nonlinear detuning increases (compare \cref{fig:detuning}).

For both cases, the initial distribution is the same before the excitation is switched on at $t=\SI{1.3}{ms}$.
The excitation band near $1-q_\x$ mainly couples to particles which perform linear betatron oscillations and are located in the core of the phase space.
When these particles are excited by the band-limited signal, they diffuse outwards in phase space towards the separatrix, causing the distribution to become flatter (\cref{fig:sim:beam_dynamics_excited}, top).
This effectively increases the particle density in the vicinity of the separatrix, which makes the process more vulnerable to fluctuations of the separatrix size caused by power supply ripples \cite{Singh.2018}.
It also intensifies fluctuations of the spill intensity caused by the diffusive nature of the noise excitation itself during the transition of particles across the separatrix.
\changed{On the other hand, the higher population of the phase space relatively close to the separatrix means, that a lower excitation strength is required to continually feed the extraction process with the desired particle rate.} %
\changed{This can be understood as an extraction driven from within the core of the phase space.}

In contrast, the excitation band near $2 q_\x$ can only couple to particles whose betatron oscillations are anharmonic and frequency modulated, since only then they have a spectral component at this frequency range.
This applies to particles towards the separatrix which undergo amplitude and phase detuning (see \cref{sec:detuning}).
As a result, the beam is \changed{extracted} from the outside and the phase space distribution becomes steeper towards the separatrix, while the particle density in the core at $H=0$ is almost unaffected (\cref{fig:sim:beam_dynamics_excited}, bottom).
\changed{In this case,} the particle density in the vicinity of the separatrix does \emph{not} increase during the slow extraction, \changed{such that} spill intensity fluctuations are effectively reduced compared to the former case.
\changed{For an equal average extraction rate, the reduced density also demands a faster flow of particles across the separatrix, which likewise reduces the sensitivity of the process to ripples.}
However, it also means that a higher excitation strength is required to make particles from the core reach and cross the separatrix \changed{in order to maintain the same extraction rate}.

\section{Conclusion}

The nonlinear particle dynamics at the \changed{third-integer} resonance used for resonant slow extraction induces a frequency modulation of betatron oscillations, which results in the appearance of additional betatron sidebands of higher order.
The second order betatron sidebands are observed prominently in \gls{btf} measurements and Schottky spectra and can be used for transverse excitation of the beam.
For \gls{rfko} slow extraction, excitation of these second order sidebands requires a larger excitation power, but improves the spill quality significantly.
As the excitation at the nonlinear second order band couples to particles near the separatrix, it effectively 
\changed{extracts the beam by pushing out particles from the outside rather than from the inside} of the phase space distribution.
Thereby, the accumulation of particles in the vicinity of the separatrix is \changed{reduced and at the same time the flow of particles across the separatrix is sped up}, which reduces the vulnerability to processes diminishing the spill quality.
\\

To improve the efficiency of the excitation process when sufficient power is not available, hybrid excitation signals can be used.
For example, a weak noise-band at the $k=1$ sideband feeding the extraction can be combined with a stronger excitation signal placed on the other side of the resonance towards the $k=2$ sideband to benefit from the described spill quality improvement mechanism.
This applies likewise to the components of more advanced excitation signals like \Npp, a combination of a band-limited noise with sinusoidal excitation signals, \changed{which is described in detail in} \cite{Niedermayer.2024a}.
In \cref{fig:exp:scan_qex_noisepp} the spill quality is shown as a function of the frequency of the noise and one of the sine components of the \Npp excitation signal used at \sis for a \ion{197}{Au}{65+} beam at \SI{200}{MeV/nucleon}.
For the noise component, frequencies between the first order sideband and the resonance are detrimental to the spill quality, and it is preferable to place the noise band at a sufficient distance to the $1-q_\x$ band or on the other side of the resonance.
For the frequency of the sine component a local optimum exists not only at $\qex[1] \approx 0.677=1-0.323$ towards the first order sideband, but also at $0.648 = 2 \times 0.324$ towards the second order sideband on the other side of the resonance.
With $1-q_\x=2/3+\abs{d}$ and $2q_\x=2/3-2\abs{d}$ the observed behaviour on the left side of the resonance is essentially a mirror image of the right side, scaled by a factor of two.

\begin{figure}
    \centering
    \includegraphics[scale=0.75]{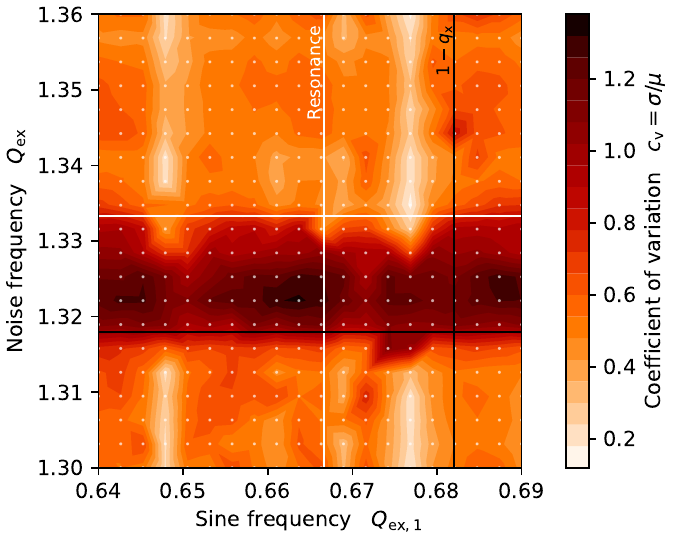}
    \caption{%
        Measured spill quality as a function of sine and noise frequency for an excitation with \Npp.
        For the scan, the noise bandwidth $\dqex=0.01$ and frequency of the second sine $\qex[2]=1.6715$ are kept constant.
    }
    \label{fig:exp:scan_qex_noisepp}
\end{figure}

In general, finding the optimal excitation frequencies is an optimization problem.
By setting appropriate boundaries, one can allow the optimization algorithm to explore also the region of the second order sideband and then chose the local optimum according to the available excitation power.
\\

\FloatBarrier

\begin{acknowledgments}

The beam instrumentation department at GSI is acknowledged for the support in carrying out the measurements.
We thank C. Cortés for valuable discussions on nonlinear BTF measurements.

\end{acknowledgments}

\bibliography{literature}

\end{document}